\documentclass[grl]{agu2001}

\usepackage{graphicx} 

\authorrunninghead{DAVIDSEN AND GOLTZ}

\titlerunninghead{Seismic waiting time distributions}

\authoraddr{J. Davidsen, Max-Planck-Institut f\"ur Physik Komplexer Systeme,
N\"othnitzer Strasse 38, 01187 Dresden, Germany.}
\begin{document}

%
%

\title{Are seismic waiting time distributions universal?}

%
%

\author{J\"orn Davidsen}
\affil{Max-Planck-Institut f\"ur Physik Komplexer Systeme,
Dresden, Germany}

\author{Christian Goltz}
\affil{Institut f\"ur Geowissenschaften, Abt. Geophysik,
Universit\"at Kiel, Germany}

%
%

\begin{abstract}
We show that seismic waiting time distributions in California and
Iceland have many features in common as, for example, a power-law
decay with exponent $\alpha \approx 1.1$ for intermediate and with
exponent $\gamma \approx 0.6$ for short waiting times. While the
transition point between these two regimes scales proportionally
with the size of the considered area, the full distribution is not
universal and depends in a non-trivial way on the geological area
under consideration and its size. This is due to the spatial
distribution of epicenters which does \emph{not} form a simple
mono-fractal. Yet, the dependence of the waiting time
distributions on the threshold magnitude seems to be universal.
\end{abstract}

%
%

\begin{article}

\section{Introduction: Scaling laws in seismicity}
\label{intro}

Earthquakes constitute an extremely complex spatio-temporal
phenomenon, with the deformation and sudden rupture of some parts
of the Earth's crust driven by convective motion in the mantle,
and the radiation of energy in the form of seismic waves. Despite
this complexity, certain characteristics of seismicity can be
captured by simple (empirical) self-similar laws. Thinking of
complexity in the sense of the physics of complex systems (e.g.
\cite{gadomski00}), one might argue that the observed scaling laws
are in fact the hallmark of the crust being a complex system (see
\cite{mulargia} for a recent review). For example, in a specified
area covering many geological faults and over a given (large) time
window the number of earthquakes $N(m>m_{th})$ with a magnitude
$m$ larger than some threshold $m_{th}$ is given by the
Gutenberg-Richter law (GR) \citep{gutenberg} which states that
$\log_{10}{N(m>m_{th})} \propto -b \; m_{th}$ with $b\approx1$
\citep{frohlich93}. Taking into account that the strain released
during an earthquake is directly related to the moment $M$ of the
earthquake (by definition $M=\mu A \delta_e$ where $\mu$ is the
shear modulus of the rock, $A$ the area of the fault break, and
$\delta_e$ is the mean displacement across the fault
\citep{turcotte}) which in turn increases exponentially with the
magnitude ($\log_{10}{M}=cm+d$ with $c=1.5$, $d=10.73$
\citep{stein}), the probability distribution of the released
strain turns out to be a power law, precisely the imprint of
scale-free behavior.

This self-similarity exhibited by GR, the independence of stress
drop on magnitude \citep{ide01,kagan02}, and recent laboratory
measurements of coseismic slip resistance \citep{ditoro04} are
good evidence that the physics of earthquake rupture is the same
for small and large earthquakes. It is still controversial if this
is different for the largest earthquakes (see \citep{kagan02} for
a discussion) because of the finite thickness of the brittle
layer.

Another example of a self-similar law in seismicity is Omori's law
\citep{omori}. It states that immediately following a main shock
there is a sequence of aftershocks whose frequency $n$ decays with
time $t$ after the main shock as $n(t)=\frac{C}{(K+t)^p}$ for
$t<t_{cutoff}$ with $p\approx1$ and fault- and magnitude-dependant
constants $K$, $C$ and $t_{cutoff}$ \citep{utsu95}. An obvious
difficulty with Omori's law is the identification of aftershocks
especially because aftershocks are not caused by a different
relaxation mechanism than main shocks
\citep{hough97,helmstetter03}. Very recently, \cite{baiesi04}
proposed a solution based on a metric measuring the correlation
between any two earthquakes.

While the existence and limitations of GR and Omori's law are well
tested and accepted, much less is known about other earthquake
statistics and their (universal) properties. Here, we will focus
on the distribution of time intervals between successive
earthquakes above a certain magnitude in a given area which is an
important quantity characterizing earthquake occurrence. In the
past, many different possibilities have been proposed for these
waiting times, from totally random to periodic occurrence of large
earthquakes. The most extended view is that of two separated
processes, one for main shocks, which ought to follow a Poisson
distribution \citep{gardner74} (or not
\citep{smalley87,sornette97,wang98}), and an independent process
to generate aftershocks. Very recently, \cite{bak02} suggested
that waiting time distributions can be described by a power-law
with a cutoff for large waiting times which scales with the size
of the considered area and magnitude threshold, indicating a
generalized scaling behavior. \cite{corral03} further proposed
that the ``cutoff'' is rather a transition between two different
power laws.

Here, we show that yet another power law regime exists for small
waiting times. Our results also provide clear evidence that the
additional transition point scales differently with area size. We
attribute this to the generic structure of the epicenters' spatial
distribution which is shown not to form a simple fractal but to
have a rather complicated structure, possibly multifractal as
proposed in \citep{hirabayashi92,goltz}.

\section{Waiting time distributions in California and Iceland}

To analyze the distribution of waiting times, we follow
\citep{bak02} and take the perspective of statistical physics by
neglecting tectonic features and any classification of earthquakes
as main shocks or aftershocks. We consider spatial areas and their
subdivision into square cells of length $L$ in kilometers. For
each of these cells, only events with magnitude above a certain
threshold $m_{th}$ are taken into account. In this way, we obtain
the waiting times $\tau_i=t_{i+1}-t_i$ in seconds between
successive events at time $t_i$ and $t_{i+1}$ in each $L^2$-cell.
Combining the waiting times from all cells, the probability
density function of the waiting times $P_{m_{th},L}(\tau)$ can be
estimated.

Analyzing data from southern California, \cite{bak02} proposed the
following scaling law
\begin{equation}
\label{ptau} P_{m_{th},L}(\tau) = \tau^{-\alpha} f\left( \tau
L^{d_f} / S^\beta\right)
\end{equation}
with $\alpha \simeq 1$, $d_f \simeq 1.2$, $\beta \simeq 1$,
$S=10^{m_{th}}$ and a scaling function $f$ depending only on the
combined argument $L^{d_f} S^{-\beta} \tau$. For small arguments,
$f$ is approximately constant not affecting the power-law
$(1/\tau)$ behavior; for large arguments, $f$ decays rapidly and
in such a way that it seems to be consistent with a different
power-law \citep{corral03}. While there are good arguments that
$\beta \equiv b$ \citep{corral03}, $\alpha$ does \emph{not}
correspond to Omori's $p$ as originally claimed by \cite{bak02}
since the decay of the rate of aftershocks determines the
corresponding \emph{all-return time} distribution but not the
waiting (or first-return) time distribution. It is an open
question if $d_f$ in Eq.~\ref{ptau} corresponds to the fractal
dimension $D_0$ of the epicenter distribution: \cite{corral03} has
found $D_0=1.6$ for southern California. It is also unclear if the
analysis by \cite{bak02} has suffered from the incompleteness of
their data for small magnitudes \citep[see][]{wiemer00}.

To test the general validity of the proposed scaling law
(Eq.~\ref{ptau}) and to clarify the scaling with $L$, we study two
earthquake catalogues --- one from southern California
($\mathcal{C}$, 19158 events) and one from southern Iceland
($\mathcal{I}$, 16286 events). For $\mathcal{C}$, the coordinates
of the polygon are $(120.5^\circ W, 115.0^\circ
W)\times(32.5^\circ N, 36.0^\circ N)$. Based on \cite{wiemer00},
the reporting of events is assumed to be homogeneous from January
1984 to December 2000 and complete at the level of $m_c=2.4$. For
$\mathcal{I}$, it was found that within $(21.43^\circ W,
19.8^\circ W)\times(63.62^\circ N, 64.30^\circ N)$ the reporting
of events was homogeneous from July 1991 to December 1995 and
complete at the level of $m_c=0.5$ \citep{wyss03}. The average
waiting time differs by a factor of $\approx 4$ from $\mathcal{I}$
to $\mathcal{C}$.

For $\mathcal{C}$, we find that indeed $P_{m_{th},L}(\tau)$ decays as
a power law with exponent $\alpha \approx 1.05$ over some range of
$\tau$. This can be deduced from Fig.~1 
and Fig.~2 
where $P_{m_{th},L}(\tau)$ is plotted in terms
of rescaled coordinates for different values of $L$. The $x$-axis
is chosen as $x = L^{d_f} S^{-\beta} \tau$ with $S = 10^{m_{th}}$,
and the $y$-axis represents $y = \tau^{\alpha}
P_{m_{th},L}(\tau)$. Thus, the constant regime in Fig.~1
and Fig.~2 
corresponds to a
power-law decay of $P_{m_{th},L}(\tau)$ with exponent $\alpha$.

Yet, Fig.~1 
and Fig.~2 
also show that
there is another, previously unnoticed power-law regime for
smaller values of $x$: The power-law increase for the rescaled
coordinates implies that the waiting time distributions decay with
exponent $\gamma \approx 0.64$. Fig.~1 
further shows
that there is a rather sharp transition between the two power-law
regimes characterized by $\alpha$ and $\gamma$, respectively.
Moreover, for $d_f=1.0$, all data collapse nicely onto a single
well-defined curve for $x<10^6$ and over 6 orders of magnitude
including the transition points. This data collapse implies in
particular that the location of the transition point $T_1$ between
the $\alpha$- and $\gamma$-regime depends on $\tau$, $m_{th}$ and
$L$ only through the combined variable $x$ with $d_f = 1.0$.
For example, keeping $m_{th}=2.4$ fixed, we find
$T_1\approx1.66$h for $L=10$km and $T_1\approx6.65$h for
$L=2.5$km. Extrapolating to $m_{th}=5$, for instance, it follows
$T_1\approx49.0$h for $L=100$km.

However, the data do not collapse onto a single curve for $x>10^6$
(see Fig.~1) 
implying that no single scaling function
$f$ exists. Thus, the proposed scaling law given in Eq.~\ref{ptau}
cannot be strictly valid. This is further confirmed by varying
$d_f$. Already for $d_f=1.2$ which was used in \cite{bak02}, the
data do not collapse onto a single curve for \emph{small values}
of $x$. For $d_f \equiv D_0 = 1.6$, the situation is even worse as
Fig.~2 
shows. Yet, with the exception of the curves
for the smallest values of $L$, $L=2.5$km and $L=5$km, the data
seem to collapse for $x>4 \times 10^3$ and over 5 orders of
magnitude. Since the estimates of $P_{m_{th},L}(\tau)$ for large
arguments and small values of $L$ might suffer from insufficient
length of the data set leading to an underestimation of the actual
$y$ values, a well-defined cutoff or transition point $T_2$
between the power-law regime characterized by $\alpha$ and the
regime for larger arguments showing a rapid decay (which might or
might not be a power law) might exist. In contrast to $T_1$, $T_2$
depends on the combined variable $x$ with a different $d_f$,
namely $d_f=1.6$, and, thus, $T_1$ and $T_2$ scale differently
with $L$. Consequently, differences in $L$ are important and
cannot be captured by a single exponent $d_f$.

For $\mathcal{I}$, we obtain similar results which are shown in
Fig.~3 
together with data from $\mathcal{C}$ for comparison: For
$d_f=1.0$ and $\beta=0.95$ and for $x<10^5$, the different curves
collapse nicely over 5 orders of magnitude despite large
variations in $L$ and $m_{th}$ and their different geographical
origin. In particular, within the statistical errors, there are no
differences between $\mathcal{I}$ and $\mathcal{C}$ in terms of
$\alpha$ as well as $\gamma$. This suggests that $\alpha \approx
1.1$ and $\gamma \approx 0.6$ are universal. The data collapse
further confirms that $P_{m_{th},L}(\tau)$ depends on $m_{th}$
only through the combined argument $S^{-\beta} \tau$ implying that
a change in $m_{th}$ leads only to a rather simple rescaling of
$\tau$. Besides, both for $\mathcal{I}$ and $\mathcal{C}$ we find
$\beta \approx b$. Thus, $P_{m_{th},L}(\tau)$ scales with $m_{th}$
and according to GR:
\begin{equation}\label{mscale}
    P_{m_{th},L}(\tau) = \tilde{P}_{L}(\tau/10^{b \; m_{th}}).
\end{equation}
Here, $\tilde{P}_{L}$ does not only depend on $L$ but also on the
specific geographical area as Fig.~3 
shows. Significant differences for large values of $x$ between the
curves with the same $L$ but from different geographical regions
are present. The waiting time distributions from $\mathcal{C}$
show a very sharp transition at and a much steeper decay for large
arguments than the ones from $\mathcal{I}$. These findings are
evidence that even a modified version of Eq.~\ref{ptau} ---
incorporating the additional exponent $\gamma$, for example ---
with a \emph{universal} function $f$ does not exist. The data from
$\mathcal{I}$, similar to those from $\mathcal{C}$, also do not
collapse for large $x$ despite identical $m_{th}$. Thus, a simple
scaling of $P_{m_{th},L}(\tau)$ with $L$ involving a single
dimension $d_f$ does generally not exist. As we show below, this
is due to the fact that the distribution of epicenters does not
form a simple fractal.

\section{Distribution of epicenters in California}

Any homogeneous (mono-)fractal is completely described by the
capacity dimension $D_0$ alone. In the case of heterogeneous
(multi-)fractals, $D_0$ describes only one, albeit dominant,
feature of the set as can be seen from the generalized definition
of fractal dimension (the spectrum of generalized dimensions
\citep{hentschel83}) $D_q = \lim_{L \to 0} \left(\frac{1}{q-1}
\frac{\ln M_q(L)}{\ln L}\right).$ Here, $M_q(L) = \sum_i P_i(L)^q$
are the generalized $q$th moments of the probabilities $P_i(L) =
N_i(L)/N$. $N_i(L)$ is the number of events found in the $i$th
cell of size $L$, and $N$ is the total number. $D_0$ is recovered
for $q=0$. Two other prominent fractal dimensions, the information
and correlation dimension, result for $q \to 1$ and $q=2$,
respectively. Higher moments increasingly emphasize densely
populated, i.e. seismically more active, areas of the set. $D_q$
may in principle be obtained for any value of $q$, especially
desirable for negative $q$ which would characterize the scaling
properties of regions of low seismic activity. Yet, due to the
relative sparsity of earthquake data and its relative inaccuracy,
we restrain our analysis to $q \ge 0$ which will nevertheless
allow us to judge the non-uniformity of the epicenter
distribution. To obtain most reliable results, we base our
numerical approach on a generalization of the correlation-integral
method as proposed by \cite{pawelzik87}.

Figure 4 
shows $D_q$ vs. $q$ for $\mathcal{C}$ and for $0 \le q \le 15$.
Inset is a double-logarithmic plot of $M_q(L)^{1/(q-1)}$ vs. $L$
for the first three generalized dimensions and for $q = 15$, the
highest moment considered. $D_q$ is estimated from the respective
slopes by fitting a straight line over the appropriate scaling
region which ranges from about 1.5 to 180 km in this case,
sufficiently large to believe in the fractal nature of the data.
The scaling region holds well even for $q = 15$. We find $D_0 =
1.60 \pm 0.13$, in agreement with \cite{corral03}. Furthermore,
$D_1 = 1.38 \pm 0.03$, $D_2 = 1.22 \pm 0.05$ and, effectively,
$D_{\infty} \approx 1$. $D_{\infty}$ is a measure for the fractal
dimension of the most densely populated vicinities in the
seismicity distribution. Most notably, $D_0$ and $D_{\infty}$ are
identical within the statistical errors to the values of $d_f$ for
$T_2$ and $T_1$, respectively. This suggests that
$P_{m_{th},L}(\tau)$ is dominated by the most densely populated
vicinities for small waiting times and the dominant feature of the
spatial distribution of epicenters captured by the fractal
dimension $D_0$ only prevails for larger waiting times. This is
exactly what is expected based on the associated difference in the
local rates of seismic activity. Taking $D_{\infty} - D_0 \approx
0.6$ as a measure of inhomogeneity, we conclude that the
seismicity distribution in $\mathcal{C}$ is definitely
heterogeneous and, thus, does not form a simple mono-fractal. For
$\mathcal{I}$, we reach the same conclusion.

\section{Discussion and conclusions}

The quality of our estimate of $P_{m_{th},L}(\tau)$ depends on
several aspects: first, on the completeness of the earthquake
catalogue. Even if a catalogue is considered to be complete above
$m_c$, certain events are missing: Directly after a large
earthquake many small events are lost in the seismic coda of the
proceeding event. Hence, short waiting times will be
underestimated and the error in $\tau_i$ can be as large as the
maximum ``deadtime'' after an earthquake. We, thus, have excluded
waiting times less than one minute.

Second, the magnitude of a given earthquake can vary as
comparisons of different definitions of magnitude show. Yet, a
detailed analysis of the data from $\mathcal{I}$ shows that
$P_{m_{th},L}(\tau)$ does not depend significantly on the
definition of magnitude chosen.

Third, while errors in the epicenter location can certainly lead
to a wrong cell association and, thus, induce errors in the
estimate of $P_{m_{th},L}(\tau)$, the (implicit) assumption that
earthquakes are a point process is much more severe. As shown by
\cite{kanamori75}, it is a good approximation to relate the moment
of an earthquake to the area of rupture by $M \propto A^{3/2}$.
Thus, $A$ increases with the magnitude of the earthquake; for
example, $A^{1/2} \approx 8 km$ for $m=6$ and $A^{1/2} \approx 80
km$ for $m=8$ \citep{turcotte}. Fortunately, for the vast majority
of earthquakes $A^{1/2}$ is significantly less than the values of
$L$ studied here. This is especially true for $\mathcal{I}$ since
the rate of occurrence of the largest earthquakes is rather small,
also preventing domination of the statistics by large events.

We are, thus, confident that our estimate of $P_{m_{th},L}(\tau)$
is reliable and the scaling regime with exponent $\gamma$ for
arguments smaller than $T_1$ is not an artifact of our analysis.
This is confirmed by the fact that the scaling of $T_1$ and $T_2$
with $L$ can be directly related to the spectrum of generalized
dimensions, namely $D_\infty$ and $D_0$, respectively. While this
directly implies that the proposed scaling law given in
Eq.~\ref{ptau} is not valid, the comparison between vastly
different scales and different geological areas suggests that
$\alpha$, $\gamma$ and $T_1$ are universal. Our results further
suggest that even if a power-law regime exists for arguments
larger than $T_2$ as proposed in \citep{corral03}, it might not
be universal. It is up to future studies if the transition from
$\gamma$ to $\alpha$ is associated with aftershock activity.

%
%

\begin{acknowledgments}
CGs work was supported by the EU project PREPARED. We thank SCEDC
and IMO for providing the data.
\end{acknowledgments}

%
%



%
%
%
\newpage 

\begin{figure}
 \label{caldf1.0}
 \noindent\includegraphics*[width=20pc]{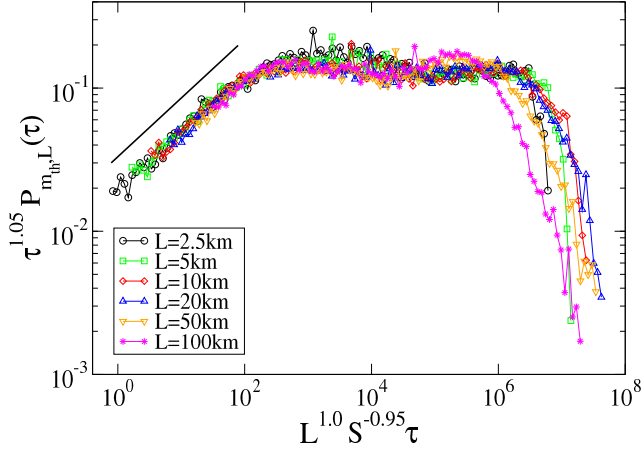}
 \caption{(color online) Rescaled waiting time distributions for California
 with $\alpha=1.05$, $d_f=1.0$, $S=10^{m_{th}}$, $m_{th}=2.4$, $\beta=0.95$.
 The solid line corresponds to a power-law decay
 of $P_{m_{th},L}(\tau)$ with exponent $\gamma=0.64$.}
\end{figure}

\begin{figure}
 \label{caldf1.6}
 \noindent\includegraphics*[width=20pc]{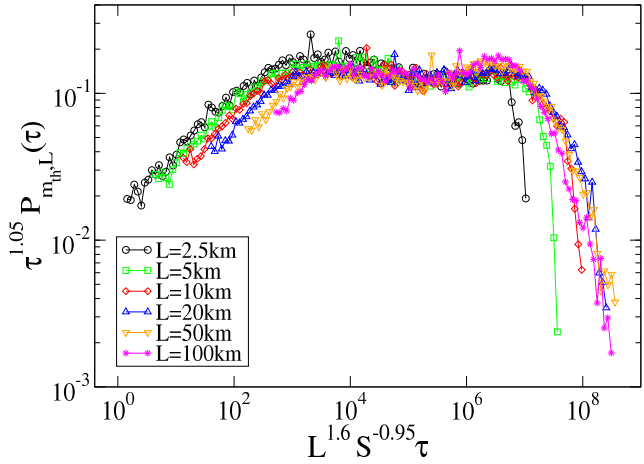}
 \caption{(color online) Rescaled waiting time distribution for California
 as in Fig.~1 but with $d_f=1.6$.
 }
\end{figure}

\begin{figure}
 \label{comb}
 \noindent\includegraphics*[width=20pc]{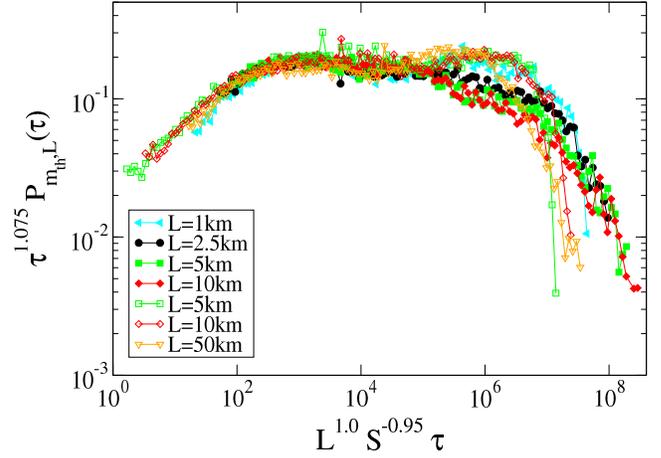}
 \caption{(color online) Combined rescaled waiting time distributions for
 Iceland (solid symbols, $m_{th}=0.5$) and California
 (open symbols, $m_{th}=2.4$)
 with $\alpha=1.075$, $\beta=0.95$, $d_f=1.0$, $S=10^{m_{th}}$.
 Since the average error in epicenter locations is
 smaller for Iceland ($<1$km) than for California ($\approx1.7$km),
 we have used different values for $L$. }
\end{figure}

\begin{figure}
 \label{d_q}
 \noindent\includegraphics*[width=20pc]{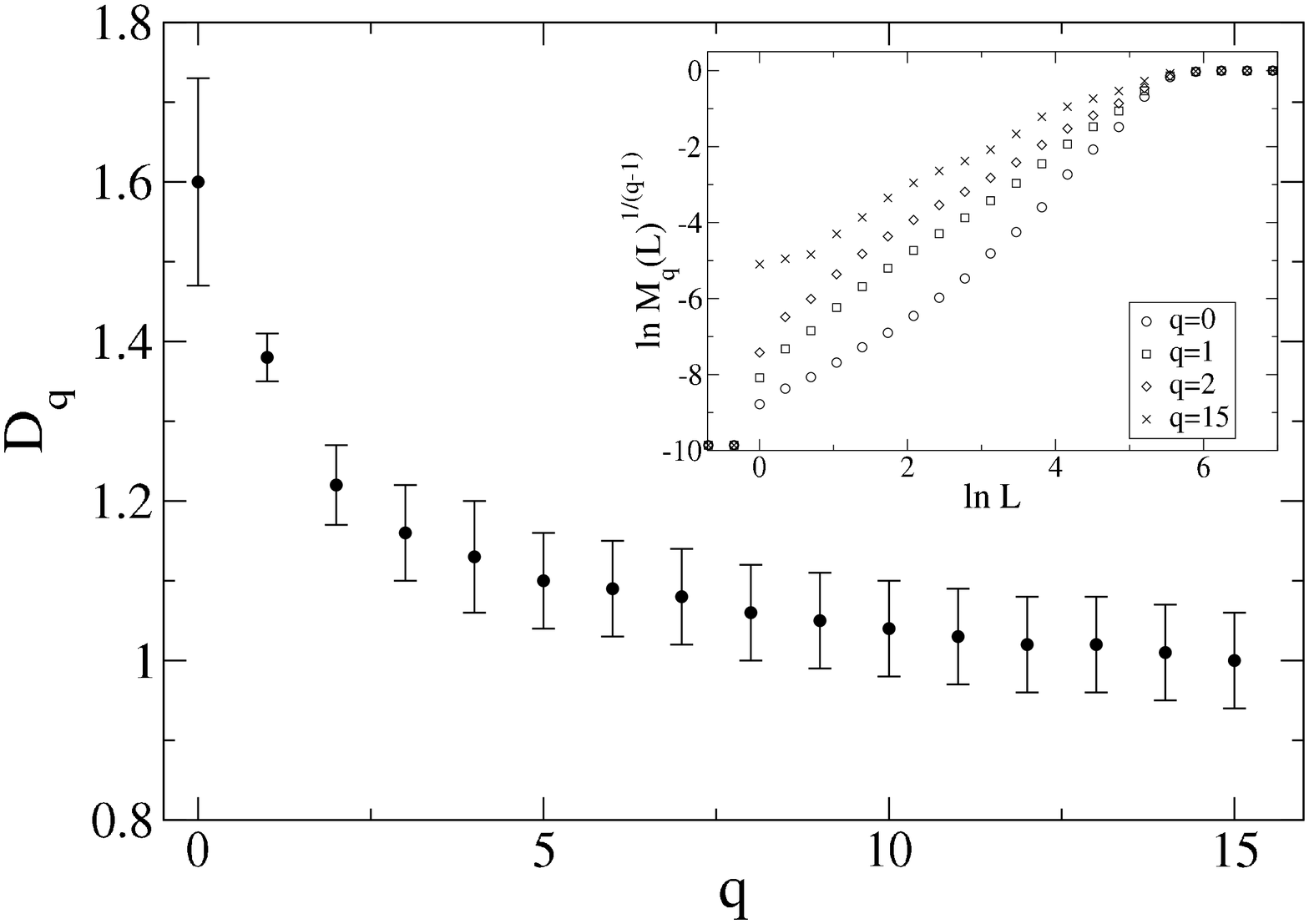}
 \caption{Spectrum of generalized dimensions $D_q$ for the epicenter
 distribution of California. Inset: generalized moments, from
 which $D_q$ is obtained.}
\end{figure}

%
%
\end{article} 

%
%
%
%

\end{document}